# The application of HEXS and HERFD XANES for accurate structural characterization of actinide nanomaterials: application to ThO$_2$.

Lucia Amidani,*[ab] Gavin B. M. Vaughan,[c] Tatiana V. Plakhova,[d] Anna Yu. Romanchuk,[d] Evgeny Gerber,[d] Roman Svetogorov,[e] Stephan Weiss,[b] Yves Joly,[f] Stephan N. Kalmykov,[d] and Kristina O. Kvashnina[ab]

[a] Dr. L. Amidani, and Dr. K. O. Kvashnina
The Rossendorf Beamline at ESRF
The European Synchrotron
CS40220, 38043 Grenoble Cedex 9, France
E-mail: lucia.amidani@esrf.fr

[b] Dr. L. Amidani, S. Weiss and Dr. K. O. Kvashnina
Institute of Resource Ecology
Helmholtz Zentrum Dresden-Rossendorf (HZDR)
PO Box 510119, 01314 Dresden

[c] Dr. G. B. M. Vaughan
ESRF – The European Synchrotron
CS40220, 38043 Grenoble Cedex 9, France

[d] Dr. T. V. Plakhova, Dr. A. Yu. Romanchuk, E. Gerber and S. N. Kalmykov
Department of Chemistry,
Lomonosov Moscow State University,
119991 Moscow, Russia

[e] Dr. R. Svetogorov
National Research Centre "Kurchatov Institute",
123182 Moscow, Russia.

[f] Dr. Y. Joly
Université Grenoble Alpes, CNRS, Grenoble INP, Institut Néel,
38042 Grenoble, France

Supporting information for this article is given via a link at the end of the document.

**Abstract:** Structural characterization of actinide nanoparticles (NPs) is of primary importance and hard to achieve, especially for non-homogeneous samples with NPs below 3 nm. By combining High Energy X-ray Scattering (HEXS) and High-Energy-Resolution Fluorescence Detected X-ray Near-Edge Structure (HERFD XANES), we characterized for the first time both short- and medium-range order of ThO$_2$ NPs obtained by chemical precipitation. With this methodology, a novel insight into the structure of NPs at different steps of their formation process is achieved. The Pair Distribution Function (PDF) reveals a high concentration of ThO$_2$ small units similar to Th hexamer clusters mixed with 1 nm ThO$_2$ NPs in the initial steps of formation. Drying the precipitates at ~150 °C promotes recrystallization of the smallest units into more thermodynamically stable ThO$_2$ NPs. HERFD XANES at Th M$_4$ edge, a direct probe of the f states, shows variations that we correlate to the break of the local symmetry around Th atoms, which most likely concerns surface atoms. Together, HEXS and HERFD are a powerful methodology to investigate actinide NPs and their formation mechanism.

## Introduction

The investigation of actinide materials at the nanoscale is emerging as a fascinating field of research, challenged by fundamental questions about their formation mechanism, their interaction with the environment, their migration capabilities, fundamental properties and chemical stability.[1,2] Despite the fact that nanotechnology has been rapidly developing since late 20th century and NPs are nowadays ubiquitous in many fields of science, the stage has been dominated by d-block systems. The f-block systems, in particular actinides, have been left behind, to the point that to date the properties of actinide materials at the nanoscale remain largely unknown. The proved tendency of actinides to aggregate in colloidal nanoparticles that are responsible for their environmental behaviour,[2] calls for an in-depth understanding of their properties as nanoclusters and nanoparticles, which can present special behaviour, reactivity and structure. Moreover, the high specific surface area of nanosized systems can find application in the design of high burn-up nuclear fuels.[3] The need for specialized facilities makes actinide research difficult and expensive. On the other hand, the increasing interest in actinides is promoting collaborations among universities, national laboratories, large-scale facilities and industries, and relevant progresses have been made. The many gaps and challenges of actinide nanoscience are recently being addressed more systematically thanks also to the increasing ability in controlling NPs synthesis.[4–11] In the roadmap to study NPs, mastering their synthesis goes hand in hand with the ability to accurately characterize the structure of the products. For actinide NPs, a field in its infancy, improvements in the structural characterization of non-homogeneous samples would enormously accelerate the understanding of the systems studied. One of the most investigated topics of radiochemistry at the nanoscale is the formation of tetravalent actinide oxide NPs in aqueous solution.[7–9,12–17] Tracking their aggregation mechanism at different chemical conditions, identifying the presence of multiple oxidation states and characterizing their surface are real challenges. Even what is considered the simpler system, $ThO_2$, for which only the tetravalent oxidation state is stable, is very debated. Th(IV) is the softest among the tetravalent actinide ions and its tendency to hydrolyse is lower compared to other An(IV). Th(IV) in solution can form not only mononuclear hydrolysis complexes, but also a number of polynuclear species.[18–23] The fluorite structure of $ThO_2$ is the ultimate product of Th(IV) hydrolysis, but its well-defined structure is often identified only after temperature treatments or as the result of ageing processes. Despite attempts to characterize Th(IV) precipitates have been made since 1960s,[24] the information on the structure and consequently on $ThO_2$ formation mechanisms in solution remains very scarce. In most cases, highly hydrolysed thorium salts form Th(IV) precipitates with ill-defined structure. In previous studies, such precipitates are classified as amorphous and in turned called "$Th(OH)_4(am)$" or hydrous oxides "$ThO_2 \cdot xH_2O(am)$" or "$ThO_2(am, hyd)$", where the amorphous character is only identified by the absence of peaks in the XRD pattern.[25–27] The short-range local structure of amorphous and crystalline Th(IV) precipitates have been investigated with EXAFS by Rothe et al.,[28] who first found that in amorphous samples the first Th – O shell is compatible with bond lengths heavily scattered around the value of crystalline $ThO_2$. Apart from the evidence of local disorder and the absence of long-range order, almost no structural information is available up to date on Th(IV) hydrous oxide.

Few works identified small crystallites of $ThO_2$ in the precipitates obtained with synthesis conditions compatible with the formation of the amorphous Th(IV) hydrous oxide.[9,29–32] Magini et al.[31] investigated hydrolysed thorium salts with wide and small angle X-ray scattering and found small clusters of atoms and microcrystalline $ThO_2$ particles up to 4 nm in heat-treated solutions at relatively mild temperatures (below 100 °C). Dzmitrowicz et al.[32] observed $ThO_2$ crystallites of more than 3 nm in X-ray amorphous precipitates using TEM and electron diffraction. Overall, the nature of Th(IV) precipitates in aqueous solution remains highly debated because of the absence of a clear-cut structural characterization of the products formed, which can be a mixture of different phases difficult to isolate. One way to solve the controversy would be to obtain monodispersed NPs, a goal that up to now was achieved using surfactants[5,33–35] or the pores of a covalent organic framework as an inert template.[14] In the first case, strong binding ligands from the organic acids alter the energetics of the surface[36] and ultimately give $ThO_2$ NPs of a given morphology and size. In the second case, Moreau et al. obtained monodispersed $ThO_2$ NPs below 3 nm and were able to structurally characterize the NPs with XANES and EXAFS. They found a fluorite structure with substantial local disorder at the surface without the need to invoke an amorphous phase.[14] In all cases, the synthesis routes use different organic Th precursors so the verification on $ThO_2$ sample produced by chemical precipitation in aqueous media is required. Ultimately, the debate around $ThO_2$ and more generally the study of actinide NPs formation need structural characterization tools able to probe both short- and medium-range order on solids and liquids.[37,38] It is indeed ideal to measure the sample without altering its state after synthesis and to characterize all relevant length scales of the system. Up to now, EXAFS is the structural technique of preference to determine anomalies in the local coordination of actinide NPs compared to bulk. However, it only provides information on the closest coordination shells.

High Energy X-ray Scattering (HEXS) and X-ray Absorption Near-Edge Structure (XANES) in the hard X-ray regime respond to these requirements and present specific advantages when applied to actinides. HEXS is among the most powerful techniques for the structural investigation of nanomaterials.[39] It measures the arrangement of atoms with ångström resolution without requiring long-range order, making it suitable for the characterization of amorphous and nanostructured systems.[40,41] HEXS is typically analysed through the pair distribution function (PDF), which is the appropriately normalized Fourier transform of the scattering signal and provides the probability to find a pair of atoms separated by a distance $r$. When applied to actinide materials, HEXS provides actinide-centric pair correlations due to the huge scattering power difference between the metal and the anion, as well as an optimal contrast with the solvent. Soderholm and co-workers were the first to make systematic use of HEXS to investigate the structure of actinide hydrolysis and condensation products and to promoted its use in actinide research.[23,37,42–44] Despite their notable results, application of HEXS to actinide systems remains limited and focused on subnano systems having only few coordination shells. To our knowledge, we provide here the first in-depth analysis of HEXS data on heterogeneous samples containing actinide NPs on the nano- and subnano scale. XANES is also very powerful for the study of nanomaterials.[45,46] The high sensitivity to the local electronic structure of a selected species is very appealing for the study of surface atoms: the sudden break of periodicity, the presence of local distortions, the rearrangement of valence charges due to dangling bonds and surfactants are all effects that affect XANES spectral shape. While in bulk materials the signal from the surface represents a negligible contribution, the surface to volume ratio increases steeply with decreasing size and in spherical NPs below 5 nm surface atoms are already few tens percent of the total amount. On such systems, XANES bears valuable information on the local structure of surface atoms. The adoption of the High-Energy-Resolution Fluorescence Detected (HERFD) mode enhances the sensitivity of XANES. The reduced core-hole lifetime broadening allows the detection of smaller spectral changes and of features that would otherwise be invisible in conventional XANES.[17,47–49] Application of HERFD to $M_{4,5}$ edges of actinide materials revolutionized the use of XANES in the field because it provided a direct probe of the 5f states with sufficient resolution to determine the oxidation state and observe the splitting due to f-electron interactions.[47,49–51] Despite $M_{4,5}$ HERFD XANES is considerably exploited in the actinide fields,[15,16,52] only very few examples applied it to NPs and to our knowledge no size-effect has been reported yet at these absorption edges.

In this work, we demonstrate the fundamental structural insight given by HEXS and HERFD XANES applied to $ThO_2$ NPs synthesized by chemical precipitation followed by thermal treatment. With HEXS, carefully analysed with model structures of NPs, we were able to distinguish and quantify particles of different sizes and in particular to detect the presence of small clusters of atoms in the first stages of synthesis. HERFD XANES spectra at the Th $M_4$ edge of different steps of the synthesis show modifications of the f density of states (DOS) which thanks to the structural insight obtained by HEXS and by using theoretical simulations, we could correlate to the break of Th local symmetry, most likely happening at the surface. The combination of these two techniques thus gives a complete view of the structure of the NPs over all relevant length scales and can tackle the structural characterization of non-homogeneous samples of actinide NPs synthesised by chemical precipitation.

## Results and Discussion

Samples of ThO$_2$ NPs were synthesized by sequential heat treatment of freshly precipitated Th(IV) samples. Sample 1 and sample 2 result from drying in air the precipitate at 40 °C and at 150 °C, respectively. To obtain ThO$_2$ nanoparticles of various sizes, the freshly precipitated Th(IV) was annealed at 400 °C, 800 °C and 1200 °C in air in a muffle furnace. According to XRD, sample 1 and sample 2 contain crystalline ThO$_2$ NPs with average coherent scattering domains of 2.0 and 3.8 nm, respectively (Figure S1). With annealing, particles grow significantly. The average size of crystallites in samples annealed at 400 °C, 800 °C and 1200 °C was around 6 nm, 34 nm and >100 nm, respectively. A table summarizing the information on samples sizes obtained by XRD and HRTEM can be found in the ESI (Table S1), while for a detailed description the reader is referred to Plakhova et al.[9]
Figure 1 shows the PDF obtained by HEXS measurements on sample 1, sample 2 and bulk ThO$_2$. All peaks in the PDF of samples correspond to peaks of bulk ThO$_2$, with the only exception of a feature of sample 1 at ~7.5 Å that will be discussed later in the text. Due to the low scattering power of O compared to Th, the signal is dominated by Th – Th and Th – O pairs. The latter appear as distinguished peaks below 7 Å, then above 7 Å the intensity drops rapidly and they become small shoulders at the bottom of Th –Th peaks. Figure S2 (ESI) shows peak assignment based on Th-centred distances in ThO$_2$ structure. Compared to bulk ThO$_2$, the signal from the samples is progressively damped with increasing $r$ and shows only moderate broadening, a direct indication of the presence of NPs. The maximum distance at which oscillations are visible, i.e. 4 nm for sample 1 and 6 nm for sample 2, marks the upper limit of NP size. Further inspection of the data also reveals that peaks of sample 1 and 2 tend to shift to higher $r$ compared to bulk ThO$_2$. This is highlighted in the upper panel of Figure 1, where data between 10 – 16 Å are superimposed and scaled.
Figure S3 (ESI) shows the relative shifts between peaks of bulk and samples in the range 0 – 20 Å. The trend of sample 2 is a linearly increasing shift to higher $r$, indicating lattice expansion[40] in agreement with what was recently reported by some of the authors based on XRD measurements.[9] The trend of sample 1 is more complex, with specific Th – Th peaks showing bigger deviations than the rest. We finally note that sample 1 presents an abrupt intensity drop after the second peak, corresponding to the first Th – Th distance, and the rest of the signal. This is not the case of sample 2, where the intensity of peaks decreases smoothly with increasing $r$.
To extract quantitative information about the size and the distribution of NPs, we first fit the data with two semi-empirical models based on imposing a size envelope function to the PDF of bulk ThO$_2$: the single sphere model, which considers the sample as an ensemble of identical spherical NPs, and the lognormal distribution of spherical NPs model. The parameters of the fits were: a scale factor, the lattice parameter $a$ and the isotropic displacement parameters ($U_{iso}$) for Th and O. In addition, the single sphere model fits the average diameter of the NPs ($P_{size}$) and the lognormal model the mean diameter ($P_{size}$) and the variance ($P_{sig}^2$) of the distribution. The parameters resulting from the fits together with the square of the residual, R$_w$, are reported in Table 1 and the comparison between fits and data in Figure 2a and b. For sample 2 (Figure 2b), both models give good fits. The average NP size from the single sphere model is 3.6 nm, while the resulting lognormal distribution spreads over a large range of sizes and is characterized by a mean size of only 0.8 nm and a variance of 0.6 nm. The latter is shown in the inset of Figure 2b, together with the envelope functions used by the semi-empirical models to modulate the signal of the bulk. Comparison of the envelope functions and visual inspection of the fits show that with the single sphere model, the signal above the average diameter, i.e. 3.6 nm, is set to zero, while with a lognormal distribution small oscillations are found also at high $r$. Indeed, the fit with the lognormal model has a slightly lower R$_w$, reflecting the better agreement with data at both low and high $r$.

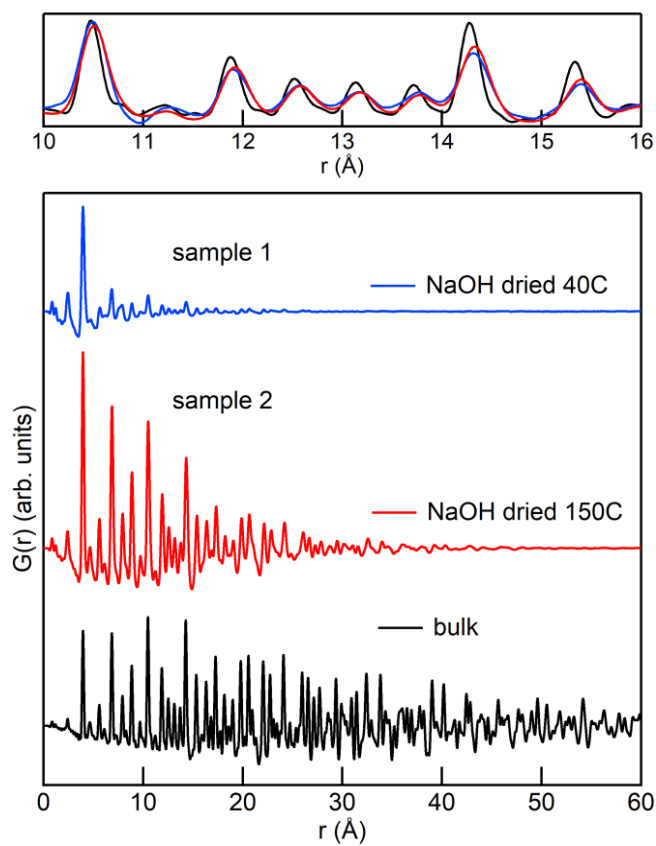

**Figure 1.** Unscaled PDF of samples 1, 2 and bulk ThO$_2$. Top panel: the same PDF data, scaled and superimposed, are shown in the 10 – 16 Å range to highlight the shift to higher $r$ of peaks of sample 1 and 2.

The results for sample 1 are shown in Figure 2a. Fitting sample 1 was more complex and we tried two $r$ ranges: 1.5 – 30 Å (fits labelled 1) and 8.3 – 30 Å (fits labelled 2). Fits 1 in the full range (Figure 2a, bottom) give poor agreement above 10 Å: both models minimize the residual at low $r$, where the signal is stronger, and they are almost featureless above 10 Å. By fitting over the full $r$ range, the single sphere model gives NPs of 0.93 nm average size and the lognormal model a very sharp distribution (P$_{sig}^2$ of 0.2 nm) peaked below 1 nm (P$_{size}$ of 0.5 nm).

**Table 1.** Fit results obtained with the semi-empirical models. $P_{size}$ is the NP size from the spherical model or the mean value of the lognormal distribution. $P_{sig}^2$ is the variance of the lognormal distribution.

|  | model | Scale | a, Å | Th $U_{iso}$ | O $U_{iso}$ | $P_{size}$, nm | $P_{sig}^2$, nm | $R_w$ |
|---|---|---|---|---|---|---|---|---|
| Sample 1 | 1 sph | 1.49 | 5.610 | 0.011 | 0.075 | 0.93 | - | 0.30 |
| | 1 logn | 1.80 | 5.607 | 0.010 | 0.080 | 0.5 | 0.2 | 0.28 |
| | 2 sph | 0.13 | 5.616 | 0.009 | 0.293 | 3.1 | - | 0.29 |
| | 2 logn | 0.18 | 5.616 | 0.008 | 0.060 | 1.5 | 0.7 | 0.31 |
| Sample 2 | sph | 0.96 | 5.619 | 0.008 | 0.059 | 3.6 | - | 0.17 |
| | logn | 1.13 | 5.619 | 0.008 | 0.063 | 0.8 | 0.6 | 0.12 |
| bulk | - | 0.57 | 5.600 | 0.004 | 0.036 | - | - | 0.11 |

$R_w$ indicates that the lognormal fit 1 is slightly better. By excluding the low *r* from the fit range as in fits 2, (Figure 2a, top), the agreement above 10 Å improves considerably. Above 10 Å the sample is well represented by uniform spheres of 3.1 nm or by a lognormal distribution peaked at 1.5 nm with a 0.7 nm variance. In contrast, when extrapolated to low *r*, fits 2 severely underestimate the signal below 5 Å. $R_w$ of fits 2, which cannot be compared with the others because of the different range, is slightly better for the single sphere model. However, visual inspection of the residuals in Figure 2a shows that fits 2 are of identical quality for the purposes of this work.

The results on sample 1 are very interesting because they indicate that the data are not well described by a single distribution or a single NP size. One characteristic size dominates the signal at low *r* and is detected by fits where the whole range is considered (fits 1). In this case, both models find average sizes below 1 nm. The residual at higher *r* definitely indicates the presence of bigger particles that can only be fitted by excluding the signal at low *r*, as done for fits 2. This is not the case for sample 2, where a single distribution is sufficient to reproduce the data. The semi-empirical models provide valuable insight into the different sizes present in the samples. However, they find considerable concentrations of NPs with diameter in the range 0.5 – 1.5 nm, predicted by assuming that diameters can take any value. This assumption is approximate below 1.5 nm and becomes appropriate only at larger diameters.

By cutting the smallest units with almost spherical shape out of a chunk of $ThO_2$ and labelling each one with the larger Th – Th distance, only few values between 0.5 and 1.5 nm are obtained. This is shown in Figure S4 of ESI. For more precise identification of the NP < 1.5 nm in our samples, we implemented a fit model based on a minimal set of $ThO_2$ NPs structures cut from the bulk. Fits were done with diffpy-CMI[53] using Debye equation.

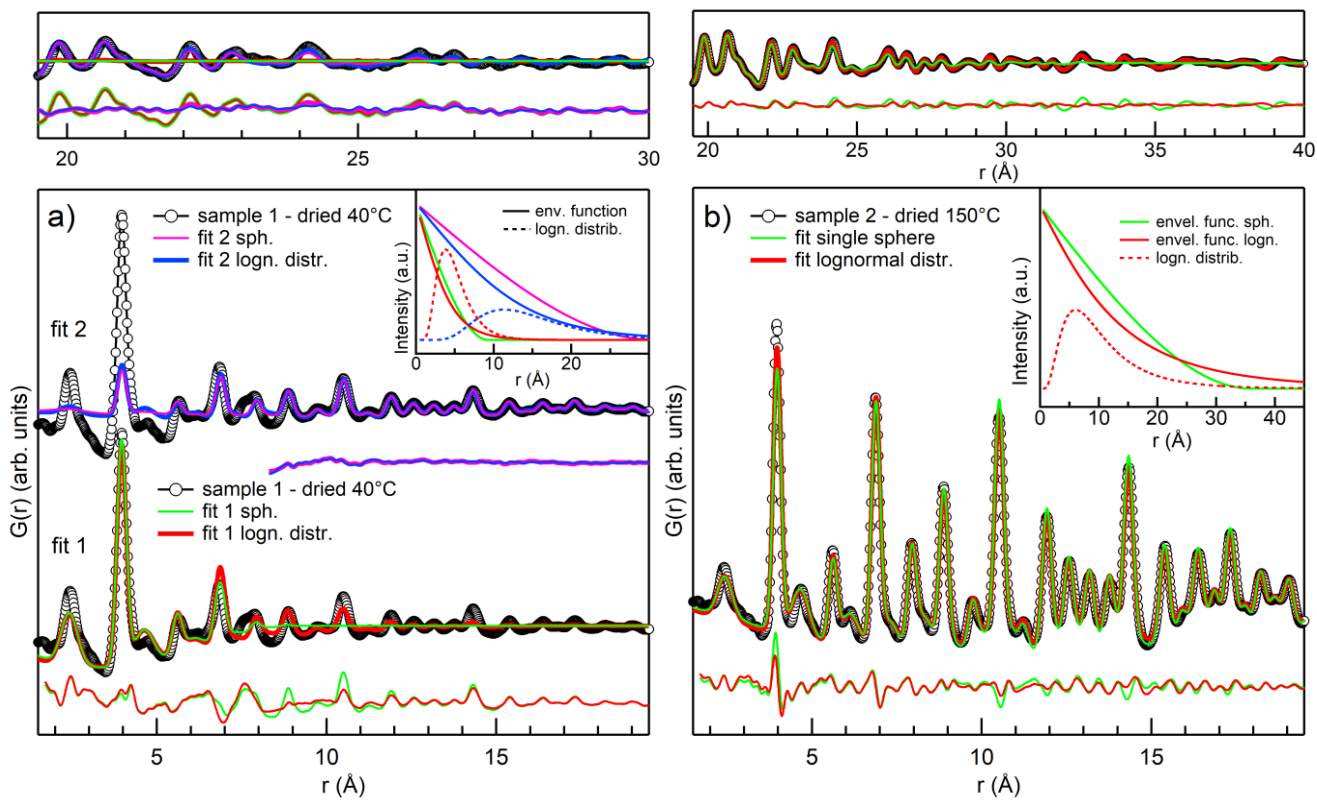

**Figure 2.** a) data on sample 1 (black circles) and fit results (coloured lines). Fit 1 is on the full range (1.5 – 30 Å), fit 2 on the reduced range (8.3 – 30 Å) but fit results have been extrapolated to 1.5 Å. Residuals are shown on the bottom of relative fits and have not been extrapolated beyond the fit range. The inset shows the envelope functions (continuous lines) for all fits together with the lognormal distributions (dashed curves). b) data on sample 2 (black circles) and fit results with the single sphere (green) and the lognormal (red) models. The inset shows envelope functions (continuous lines) for both fits and the resulting lognormal distribution (dashed line). The top panel of a) and b) shows the high *r* range.

# FULL PAPER

**Table 2.** Fit results for sample 1 and 2 with the set of NP structures.

| NP | Scale | concentration | $a^*$exp. coeff. | Th $U_{iso}$ | O $U_{iso}$ | $R_w$ |
|---|---|---|---|---|---|---|
| *Sample 1 – dried 40 °C* | | | | | | |
| 0.56 nm | 0.60 | 61.3% | 5.614 | | | |
| 1.0 nm | 0.24 | 24.5% | 5.580 | 0.0133 | 0.0467 | 0.21 |
| 2.0 nm | 0.072 | 7.4% | 5.601 | | | |
| 3.5 nm | 0.067 | 6.8% | 5.619 | | | |
| *Sample 2 – dried 150 °C* | | | | | | |
| 1.0 nm | 0.24 | 24% | 5.603 | | | |
| 2.5 nm | 0.41 | 41% | 5.603 | 0.0075 | 0.0075 | 0.09 |
| 5.6 nm | 0.34 | 34% | 5.624 | | | |

We isolated from bulk $ThO_2$ a set of NPs of almost spherical shape with diameters between 0.5 and 6.0 nm. We carefully cut all structures below 1.5 nm and above 1.5 nm we constructed spheres centered on Th with increasing radius up to 5.6 nm. The list of structures considered is reported in ESI. We fitted samples 1 and 2 in the ranges 1.7 – 40 Å and 1.7 – 60 Å, respectively, with the minimal subset of structures. PDF of ideal structures are calculated from the atomic coordinates using the Debye scattering equation implemented in diffpy-CMI (DebyePDFGenerator and DebyePDFCalculator). Each NP structure adds to the fit two parameters: a lattice expansion coefficient and a scale factor. The latter, when divided by the sum of all scale factors, gives the concentration of the corresponding structure in the sample. The isotropic displacement parameters ($U_{iso}$) for Th and O common to all structures were also fitted. In order to find the best fit, we first added big NPs and optimized the agreement in the tail of the PDF signal, where oscillations are weak and only biggest NPs contribute. Extrapolating the fit to lower $r$ and comparing it with data reveals where intensity is still missing and allows estimating which sizes to include in the ensemble to improve the fit. Due to the limited number of structures and samples, we proceeded with a manual fit that allows visual inspection of results. The results of the new fits are reported in Table 2. Figure 3a and b report the new results in comparison with those of the lognormal fits from semi-empirical models. The new fits improve the agreement for both samples. For sample 2 the $R_w$ decreases slightly and the inspection of the residuals in Figure 3b reveals small improvements over the full $r$ range. For sample 1, the $R_w$ improves considerably compared to that of lognormal fit 1, which was over the full range. $R_w$ of lognormal fit 2 cannot be compared since a different range was used. However, the direct comparison shown in Figure 3a illustrates that above 6 Å the fits give very similar results while below 6 Å the new fit reproduces very well the abrupt drop of intensity. According to the results reported in Table 2, sample 1 is made for 61.3% of 0.56 nm NPs, i.e. the smallest units that can be cut from bulk $ThO_2$, mixed with 24.5% of 1.0 nm NPs and small concentrations of 2.0 and 3.5 nm NPs. Sample 2 is made by a more homogeneous mixture of 1.0 nm (24%), 2.5 nm (41%) and 5.6 nm (34%) NPs.

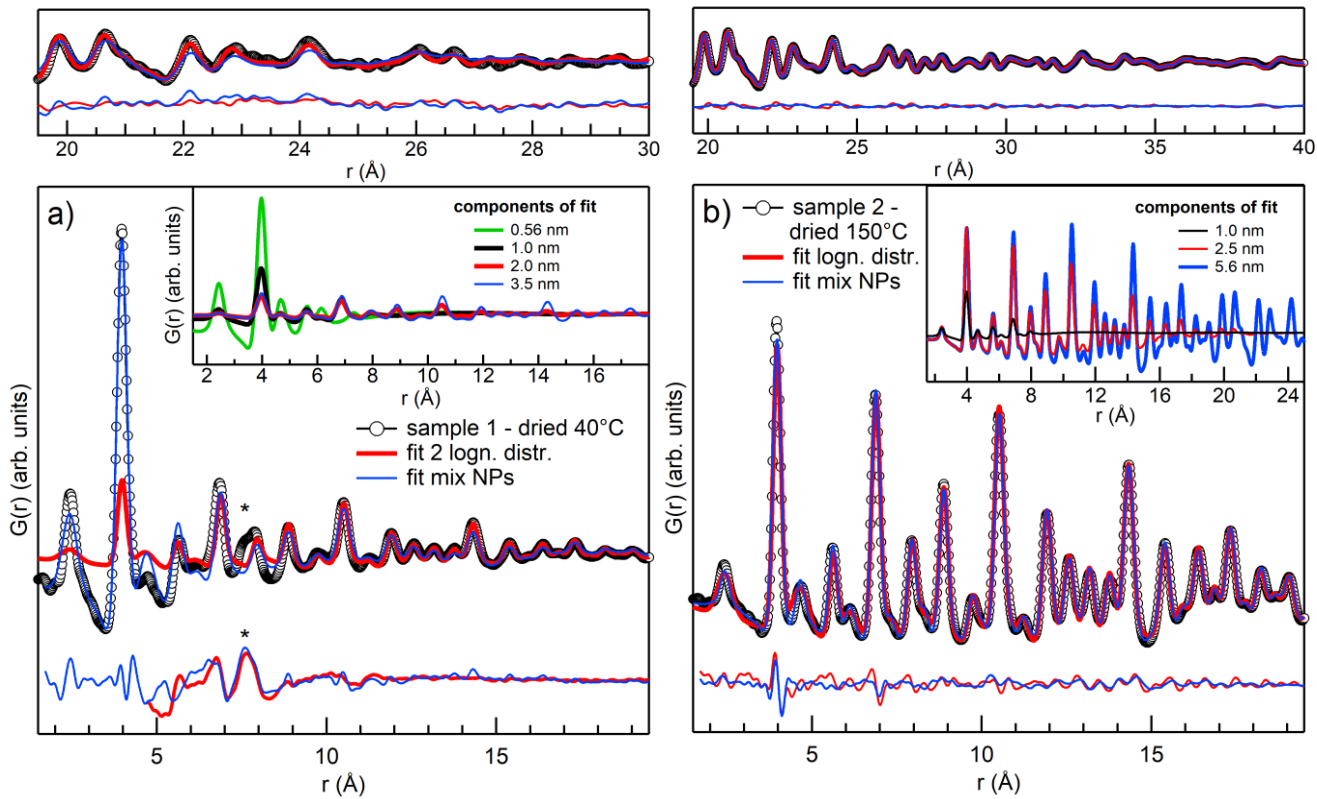

**Figure 3.** a) data on sample 1 (black circles) compared to fit results for the lognormal fit 2 (red line) and the fit with NP structures (blue line). b) data on sample 2 (black circles) compared to fit results for the lognormal fit (red line) and the fit with NPs structures (blue line). Residuals are shown below the fits and the insets of a) and b) show the calculated PDF for each NP structure added in the fit. The high $r$ range is shown in the upper panels.

The calculated PDF from each NP contributing to the fits are shown in the insets of Figure 3a and b. These results confirm what suggested by the semi-empirical models: sample 1 has a high concentration of very small particles mixed with bigger ones, while sample 2 is described by a more homogeneous distribution of sizes. Notably, the high concentration of 0.56 nm units disappears upon heating at 150 °C. This small octahedral unit that we artificially cut from $ThO_2$ bulk is very similar to Th hexamer clusters, which have been frequently reported in literature.[18–21,44] We notice that even if we applied a fit exclusively based on NP structures, a fit approach mixing NP structures for small sizes and a lognormal distribution gives results of very similar quality. Figure S5 and Table S2 in ESI report the comparison of these two approaches for sample 1. Nevertheless, with the fit using only NP structures a lattice parameter for each structure can be fit, different NP shapes can be easily implemented as well as core-shell structures. The high flexibility of this method can help to get even more detailed structural information and could be exploited in studies on larger data sets.

Despite the good results obtained, there is still room for substantial improvement, especially for sample 1. The main contribution to the residual is the peak at 7.6 Å, indicated with an asterisk in Figure 3a, which does not belong to fluorite $ThO_2$. A similar feature was previously reported by Magini et al.[31] who investigated hydrolysed Th salts. They assigned it to aggregates of O-centred tetrahedra sharing facets, an early stage of the synthesis of $ThO_2$, which is made by a network of O-centred tetrahedra sharing edges. In our case, the peak could be the result of surface disorder on 1.0 nm NPs. The disorder could cause a splitting of the peak at 7.75 Å, which would correspond to Th at the surface. The assignment of this peak, as well as a deeper insight into the early stages of Th(IV) hydrolysis, requires the collection of a bigger set of data which will be the focus of future investigations.

Figure 4 shows Th $M_4$ edge HERFD data collected on samples 1 and 2 and on samples annealed at 400 °C, 800 °C and 1200 °C. XANES at the $M_4$ edge of actinides corresponds to the excitation of an electron from the $3d_{3/2}$ to the $5f_{5/2}$ and provides direct access to the f-DOS of the actinide. The spectrum of 1200 °C annealed sample is identical to that of bulk $ThO_2$ and presents four main features: the main peak A at the absorption edge, two shoulders labelled B and C, and feature D well separated by the absorption edge region. Features C and D are absent in sample 1 and only slightly visible in sample 2 and they progressively grow for NPs annealed at high temperatures. We also note that feature B is slightly higher in sample 1. However, the difference is very small and cannot be associated with a trend like for features C and D. Further investigation on larger sets of samples is needed to confirm the effect on feature B.

The progressive growth of features C and D in Th $M_4$ edge HERFD seems to follow the increase of crystallinity and size of NPs. The results of PDF fitting indicate that the precipitate dried at 40 °C (sample 1) is predominantly made of small units similar to Th hexamers and 1.0 nm NPs. Drying the precipitate at higher temperature (sample 2) already stimulates the growth of the existing NPs and causes the extinction of the smaller units detected by PDF. From XRD data, we know that the growth continues with annealing at 400 – 1200 °C. The sensitivity to crystallinity and NPs size at the Th $M_4$ edge is quite a novelty and it may sound surprising. Indeed, 5f states are generally considered strongly localized, not involved in chemical bonding and only mildly sensitive to the crystal field of neighbouring atoms. This description fits better 4f states of lanthanides rather than 5f states of the early actinides. The latter are spatially more extended, more sensitive to the presence of neighbouring atoms and more prone to participate in bonds. The case of actinyl ions, where the actinide forms two very short and strong linear bonds with oxygen atoms, is a well-known case of chemical bond involving 5f orbitals.[54]

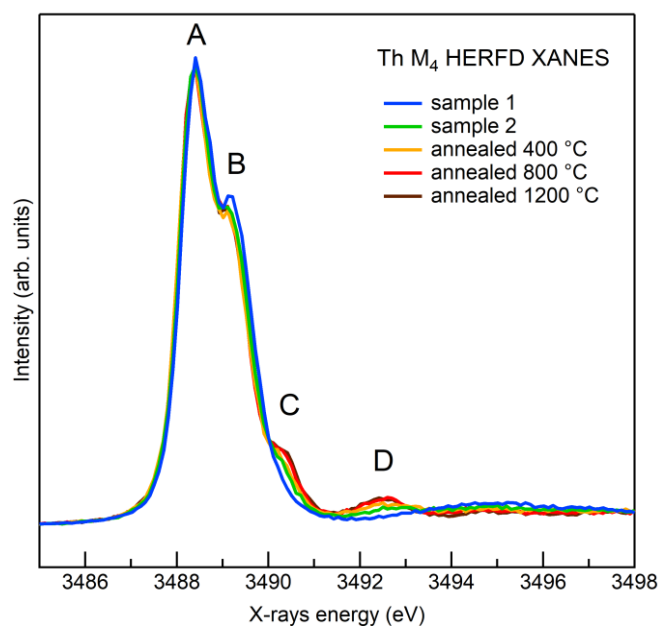

**Figure 4.** Th $M_4$ edge HERFD data on sample 1, 2 and samples annealed at 400 °C, 800 °C and 1200 °C. Data were normalized to the total spectral area.

To confirm if the observed effects can correlate with size reduction, we need to understand the nature of each feature in the spectrum and rationalize if the disappearance of peaks C and D is compatible with this hypothesis. In the absence of a large set of well-characterized references, simulations are the only way to shed light on the nature of spectral features. Butorin et al.[55] recently modelled the $M_4$ HERFD of $ThO_2$ within the Single Impurity Anderson Model, which fully accounts for electron correlations and treats the inter-atomic interactions as a perturbation. Features A and B were well reproduced by the $O_h$ crystal field effect on 5f orbitals of Th, while feature D is obtained by adding the ligand-to-metal charge transfer driven by Th 6d – Th 5f – O 2p hybridization. Even if some multiplet poles arise in correspondence of feature C, they are too weak to generate a shoulder and the assignment of feature C remains open. Within the approach used by the authors, based on atomic physics, it is difficult to implement effects due to size and to complex local distortions, because the influence of neighbouring atoms is included as a perturbation whose strength is regulated by empirical parameters. The number of parameters increases very fast with the lowering of local symmetry, like that expected for surface atoms or small clusters. Approaches that naturally account for the surrounding atoms, like those based on density functional theory (DFT), have the advantage of avoiding empirical parameters to account for local symmetry and redistribution of valence change. These approaches include electron correlations only partially, making them not suited to treat strongly correlated f-systems. Indeed, this theory fails to reproduce the $M_{4,5}$ edges of 4f elements (lanthanides). The 5f states are less localized and the calculations done following this scheme are less questionable if the purpose is to reproduce observed trends and to deduce valuable information. Moreover, for actinide materials as $ThO_2$, uranyl-type or U(VI) compounds which have empty 5f orbitals in the ground state, the DFT approaches are well-suited. To our knowledge, only a few attempts have been made to simulate XANES at $M_{4,5}$ edges of early actinides using DFT-based codes.[56–58] The outcomes were promising, even if the implications of disregarding electron correlations in systems with f-electrons have not been discussed explicitly and a comparative study on $M_{4,5}$ edges of early actinides simulated with the two approaches is still missing.

We simulated the f-DOS and the $M_4$ edge XANES of bulk $ThO_2$ with FDMNES[59,60] to elucidate the nature of spectral features. Figure 5 reports the f-DOS with both core-hole and spin-orbit effects included (panel a), with only the core-hole (panel b) and without both (panel c). Each panel reports the total f-DOS (black curve) and the Crystal Overlap Orbital Populations (COOP) between the 5f orbitals of Th and the 2s (black dot-line) and 2p (red dot-line) orbitals of neighbouring O. COOP quantify the covalency of the Th – O bond by integrating the product of their atomic orbitals inside a sphere centred on the bond axis.[61] Positive/negative COOP indicate bonding/anti-bonding character. Panel b and c of Figure 5 report also the decomposition of the total f-DOS on the cubic set of f-orbitals, which in $O_h$ symmetry splits into three groups, sketched in panel c. We first notice by comparing panel c and b that the core-hole has a strong impact on the f-DOS. It pulls it down in energy and it increases its sharpness and intensity close to the edge. Comparison of panel a and b shows that the effect of spin-orbit is to introduce additional splitting to that originating from local geometry. The final f-DOS resulting from the inclusion of the core-hole and the spin-orbit effects (panel a, black curve) has three groups of features whose energy separation and relative intensity are in good agreement with the experimental data. For further insight into the nature of each group of features, we can look at the decomposition of the total f-DOS on the cubic set for simulations without spin-orbit (panel b and c). $ThO_2$ crystallizes in the fluorite structure (Fm-3m space group) where Th has eight neighbouring oxygen atoms placed in the vertices of a regular cube. Under this local symmetry, the f-orbitals expressed in the cubic set split in three groups: $T_{1u}$ ($f_{x^3}$, $f_{y^3}$, $f_{z^3}$), $T_{2u}$ ($f_{x(y^2-z^2)}$, $f_{y(z^2-x^2)}$, $f_{z(x^2-y^2)}$) and $A_{2u}$ ($f_{xyz}$). This is indeed what we observe in Figure 5, panel b and c. The ordering in absence of a core-hole is in agreement with expectations and with previously published results for $UO_2$:[62] the contribution at lowest energy is of $T_{1u}$ symmetry (small blue peak at ~6.8 eV) while the single orbital $A_{2u}$ ($f_{xyz}$) is the more destabilized and highest in energy. Indeed, $f_{xyz}$ points towards the vertices of the cube where the O atoms are and it interacts more strongly with them. This is also seen in the COOP, which find strong bonding interaction with O 2s in correspondence of $A_{2u}$. It is interesting to notice that the core-hole (panel b) pulls all orbitals down and confers $T_{1u}$ and $T_{2u}$ extreme sharpness and high intensity, suggesting an increased localization of these orbitals in presence of the core-hole potential, while $A_{2u}$ remains isolated at higher energy and is broadened. The bonding character of Th 5f $A_{2u}$ – O 2s is maintained with the addition of significant anti-bonding interaction with O 2p.

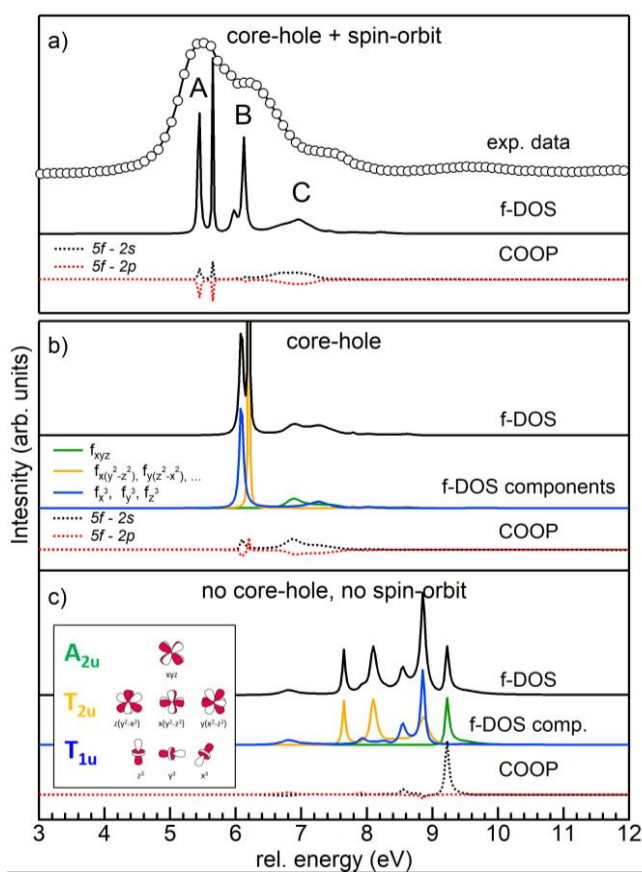

**Figure 5.** FDMNES simulations of bulk $ThO_2$ compared with experimental data on 1200 °C annealed sample (panel a – black circles). Panel a) shows the simulation obtained with core-hole and spin-orbit included: the total f-DOS (black line) and the COOP of Th 5f with O 2s (black dot line) and 2p (red dot line). Panel b) and c) report the results obtained without spin-orbit and with (b) or without (c) core-hole. The total f-DOS and the COOP are reported with the same colour code of panel a). In addition, in panel b) and c) the decomposition on the f-DOS on the cubic set is shown ($A_{2u}$ in green, $T_{2u}$ in yellow and $T_{1u}$ in blue). A sketch of the f-orbital cubic set in $O_h$ symmetry is drawn in panel c.

Unfortunately, we cannot decompose the f-DOS of panel a, obtained with spin-orbit, on a basis set symmetrized for the double group $O_h^*$. However, we know from group theory that $T_{1u}$ and $T_{2u}$ are further split in two subgroups and $A_{2u}$ remains unaffected. The parallel with the results of panel b suggests the assignment of peak A and B of the experimental $M_4$ HERFD to f-orbitals of $T_{1u}$ and $T_{2u}$ symmetry and peak C to the single $A_{2u}$ orbital. COOP shown in panel a are also very similar to those of panel b and further support our assignment. Despite the observation that the f-DOSs calculated by FDMNES are in good agreement with experimental data, the simulated XANES (see Figure 6d) calculated once spin-orbit is added completely disregards part of the f-DOS and peak B is not reproduced. The difference between the f-DOS and the simulated XANES stems from selection rules, which allow transitions only to states with j=5/2 for $M_4$ while transitions to both j=5/2 and j=7/2 states are allowed for $M_5$.[63] Figure S6 in ESI shows the comparison between simulations of $M_4$ and $M_5$. The collection of $M_5$ data would help clarifying the absence of peak B in the simulations. Unfortunately, with the spectrometer used in this experiment, a Johan-type spectrometer operating at 65 – 90°, the emission energy needed for Th $M_5$

HERFD is not covered. Measurements of Th $M_4$ spectra are only feasible with spectrometers having a different design.[64]

The total f-DOS splitting grasps the physics behind spectral features since the energy separation and relative intensity of the f-DOS agree fairly well with experimental data. We can then conclude that peaks A, B and C arise from the combined splitting of the crystal field and the spin-orbit effect on Th 5f orbitals. Peak C, in particular, is the most affected by the interaction with oxygen nearest neighbours. As expected, we do not reproduce peak D, which arises from ligand to metal charge transfer,[55] an effect that the current approach disregards.

We know from PDF that the principal constituents of sample 1 are small units similar to Th hexamer clusters and $ThO_2$ NPs of 1.0 nm. Both structures have few Th atoms (6 and 13) that, except for the central Th in the 1.0 nm NP, are on the surface and most probably experience a variation of their local environment. The disappearance of peak C in the XANES of sample 1 can then reflect this change. To confirm this hypothesis, we simulated Th $M_4$ XANES of two Th hexamers whose structure are reported in literature[18,20] and of the 0.56 nm $ThO_2$ NP that we used in PDF fitting. Both Th hexamers are made of 6 Th atoms linked together as in $ThO_2$. We added 6 $H_2O$ molecules to the 0.56 nm unit similar to what is found in Th hexamers. Figure 6 shows schematics of all the structures and compares the f-DOS and the XANES simulations of Th clusters and bulk $ThO_2$. All simulations were shifted in energy to have the first feature aligned. Interestingly, the total f-DOS of the 0.56 nm NP maintains the three features observed in the bulk, while Th hexamers only have the first two intense features and lost the last, corresponding to peak C. The XANES simulations follow the f-DOS results, exception made for the systematic absence of the second peak, which confirms that selection rules are the origin of the discrepancy. While Th hexamers (Figure 6a and b) reproduce the observed disappearance of peak C, the very similar 0.56 nm NP does not (Figure 6c). The reason is the orientation of the four external O bound to each Th: in the 0.56 nm NP they sit on the corner of the cube centred on Th, as in bulk $ThO_2$. In Th hexamers, the same O atoms are rotated of ~45° around the axis connecting two opposite Th. This rotation breaks the $O_h$ symmetry around Th and stabilizes more the anti-bonding $A_{2u}$ orbital whose superposition with O is lowered. We additionally remark that the relative intensity of the f-DOS peaks of Th hexamer differs from that of bulk $ThO_2$. In particular, the features corresponding to peak B are more intense in Th hexamer, a hint that the relative intensity of peak A and B can also be sensitive to changes of the local environment and that the slight increase of peak B observed for sample 1 may be relevant.

HEXS and HERFD data when thoroughly analysed as we did in this work, demonstrate to be a very powerful tool to unravel the complexity of $ThO_2$ NPs structure. PDF reveals that samples are a mixture of particles of different sizes and that a large percentage of subnano NPs is initially formed, which then disappears to favour bigger NPs when Th(IV) precipitate is dried at 150 °C. This finding suggests a formation mechanism where small units, which we found to be similar to Th hexamers, are incorporated into bigger NPs or aggregate with each other. The strongest evidence of the presence of subnano units comes from PDF sensitivity to both short- and medium-range order: the drop of intensity of Th – Th peaks above 5 Å can only be reproduced by units with only two Th – Th distances. We notice that at these distances the presence of additional Th atoms whose position deviates from those of crystalline $ThO_2$, would still contribute to the PDF signal, differently from what would happen for EXAFS. The latter has a higher resolution and is only sensitive to short-range order. As a result, a spread of bond distances that does not suppress the coherence needed to observe PDF peaks is sufficient to flatten the signal from higher shells in EXAFS. Figure 7 shows the simulated PDF signal for the 0.56 nm unit cut from $ThO_2$ and for the Th hexamer from.[20] On the bottom of Figure 7, all Th-centred distances are reported as sticks and show the local disorder present in Th hexamer. Despite for Th hexamer the spread of Th – O distances covers a range between 2.37 – 2.62 Å, the first peak of the simulated PDF is not affected dramatically, while a similar spread has a strong impact on EXAFS as reported by Rothe et al.[28] The few structural information on amorphous hydrous $ThO_2$ come indeed from EXAFS measurements and are limited to the first Th – O shell, which can only be fitted with multiple Th – O distances differing from those of crystalline $ThO_2$. The signal from higher shells is typically very low making impossible to disentangle and quantify static disorder and coordination numbers. The sensitivity of HEXS to short- and medium-range order allows to characterize the sample on all the relevant length scale and reveals the presence of subnano units mixed with bigger NPs. NPs below 1.5 nm have significant local disorder, but they still preserve the $ThO_2$ structure at short- and medium-range. Our study suggests that subnano units similar to Th hexamer clusters rather than an amorphous phase are the intermediate step leading to $ThO_2$ NPs formation. This picture is in line with the findings by Hu et al.[44] and with the more general emerging understanding that tetravalent-cations hydrolysis can result in well-defined metal oxide/hydroxide aggregates.[11,23]

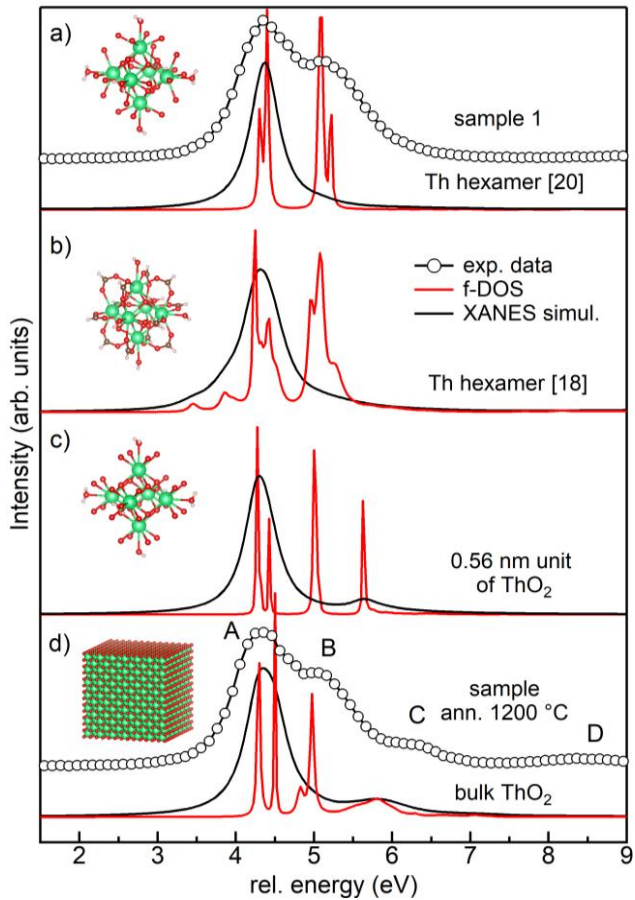

**Figure 6.** Simulations of the total f-DOSs (black lines) and XANES simulations (red lines) for Th hexamers from [20] (a) and [18] (b), the 0.56 unit from $ThO_2$ (c) and bulk $ThO_2$ (d). Experimental data on sample 1 and on 1200 °C annealed sample are shown on top of simulations in a) and d), respectively. Each panel reports a scheme of the simulated structure.

Differently from HEXS, XANES is sensitive to the short-range order and to the stereochemistry around the absorber. More details about the local coordination of the absorber can be extracted. The complexity of XANES analysis is often a bottleneck to extract the information. However, theoretical simulations available today can greatly help their interpretation and knowledge of the structure of the sample can guide the attempts to theoretically reproduce the spectral changes. In this regards, the coupling of HEXS and XANES provides more structural information and when applied to small objects can be an invaluable tool to investigate surface modifications. In the present study, we demonstrated that the trend observed in Th $M_4$ HERFD XANES can be rationalized as a change of local symmetry around Th atoms as the one found in Th hexamer clusters, where the rupture of the $O_h$ symmetry is caused by the rotation of O ligands. Such break of local symmetry is expected to affect in particular surface atoms, where dangling bonds will leave more freedom for rearrangement. The predominance of surface atoms in sample 1, whose main constituent are NPs < 1.5 nm, results in the disappearance of feature C and D of $M_4$ HERFD, which are specifically sensitive to oxygen neighbours and to Th 6d – O 2p hybridization,[55] respectively. Compared to our previous results on analogous samples,[9,17] this investigation determines with more accuracy the characteristic sizes composing the samples. In particular, the presence of NPs < 1.5 nm is in agreement with the very low Th – Th coordination number found by EXAFS and with the effect previously observed at Th $L_3$ edge XANES and ascribed to low coordinated Th atoms at the surface and disorder at the surface.[17]

By illustrating the opportunities of using HEXS and HERFD XANES to investigate $ThO_2$ NPs, our work opens the way to a thorough investigation of the mechanism of NP formation for $ThO_2$ and more generally for actinide oxides. The effects of initial chemical conditions, polymerization of the initial Th(IV) precipitates and the stability over time of the NPs can be addressed with this methodology. The latter is of particular interest since $ThO_2$ NPs, differently from NPs of other actinide dioxides synthesized with the same chemistry route, change with time (ageing effect), increasing size and crystallinity depending on the conditions. Our findings on $ThO_2$, compared with recent results on $CeO_2$[65] and $PuO_2$[15,16] NPs synthesized under similar conditions, already show differences on systems which are often indicated as very similar. The fact that Th can take only one oxidation state, while Ce and Pu not, prevent the accommodation of charge unbalance with oxidation or reduction at selected sites. This limitation may be compensated by increased disorder.[9,14,17] Comparative studies of the structural properties of $AnO_2$ NPs can help to assess similarities and differences and shed light on the different nucleation mechanisms. Altogether, the data and the analysis presented demonstrate that investigating very small NPs with HEXS and HERFD provides important insight into the structure of NPs and the local environment at their surface. Combining the two techniques is particularly important when samples are far from being ideal, i.e. with very narrow size distribution and a uniform shape. Being able to extract structural information on such samples is of paramount importance because it sheds light on mechanisms of NPs formation that may be acting in the environment.

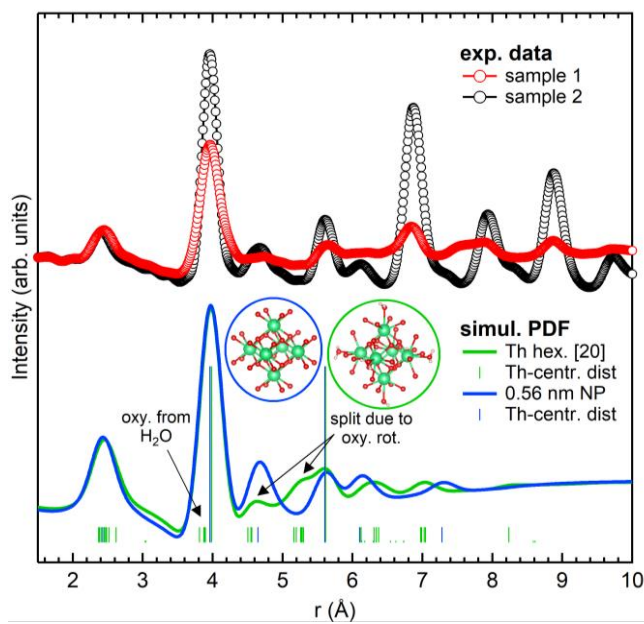

**Figure 7.** PDF experimental data of sample 1 (red circles) and 2 (black circles) are shown together with the calculated PDF for the structure of the 0.56 small unit cut from bulk $ThO_2$ (blues line) and Th hexamer from [20] (green line). Vertical sticks on the bottom indicate the Th-centred pair distances in the two structures. The height of the sticks is proportional to the atomic number (Th = 90, O = 8, H = 1) and not to the frequency of the represented distance.

## Conclusion

We investigated the structure of $ThO_2$ NPs in two samples representing the initial and a more advanced step of their synthesis via chemical precipitation. We did it by combining HEXS and HERFD XANES at the $M_4$ edge of Th. The analysis of the PDF with semi-empirical methods and with fits based on real NPs structures revealed that samples dried at mild temperature (40 °C – sample 1) contain mainly $ThO_2$ NPs < 1.5 nm. Drying at higher temperature (150 °C – sample 2) results in bigger NPs and no subnano units. HEXS reveals that Th(IV) initial precipitate contains a large number of seeds for $ThO_2$ NPs. Moreover, increasing the drying temperature to 150 °C promotes recrystallization into the more thermodynamically stable $ThO_2$ phase. HERFD at Th $M_4$ edge shows unexpectedly a remarkable effect on two spectral features, i.e. peak C and D, which progressively grow with the increased crystallinity and particles size. FDMNES simulations shed light on the nature of feature C and revealed its marked sensitivity to the break of $O_h$ symmetry around Th ions, which considered the morphology of our samples is most likely caused by the modified arrangement of O ligands at the surface. Th $M_4$ edge HERFD demonstrates a high sensitivity to the local structure at the surface when applied to small NPs, a piece of information very difficult to obtain and of fundamental importance for the understanding of interaction mechanisms of NPs. HEXS and HERFD supported by a suitable analysis and theoretical simulations are a powerful tool to investigate actinide nanomaterials, especially when non-homogeneous samples like those obtained in solution chemistry are the focus of the investigation.

## Experimental Section

ThO$_2$ samples were synthesized by sequential heat treatment of freshly precipitated Th(IV) samples. We first mixed aqueous solutions of 0.1 M thorium nitrate pentahydrate and 3 M sodium hydroxide, which results in the formation of a Th(IV) precipitates. Portions of the freshly precipitated Th(IV) sample were dried in air at 40 °C (sample 1) and at 150 °C (sample 2). To obtain ThO$_2$ nanoparticles of various sizes, the freshly precipitated Th(IV) was annealed at 400 °C, 800 °C and 1200 °C in air in a muffle furnace. A summary of sample synthesis and their structural characterization by XRD and HRTEM can be found in Electronic Supplementary Information (ESI) (Table S1 and Figure S1).

We collected HEXS on sample 1 and sample 2 at the ID15A[66] beamline of the European Synchrotron Radiation Facility (ESRF, Grenoble). The data were collected at room temperature, the incident energy was set to 120 keV and we measured up to 30 Å$^{-1}$ using a Dectris CdTe 2M pixel detector. Samples were sealed in kapton capillaries (double confinement) and the signal from an empty capillary was used for background subtraction. Patterns were corrected for detector geometry, response and transparency, and integrated using a locally modified version of pyFAI[67] with outlier filtering.

The PDF was calculated from the resulting powder diffraction patterns using modules from PDFgetX3[68] and The data were corrected for electronic noise and weak spurious signal by fitting the high angle part of the calculated F($q$) to a weighted spline in order to remove outliers, in a procedure similar to that described in reference [69]. The PDF Gaussian dampening envelope due to limited Q-resolution and the Q-broadening were obtained from the fit of a reference sample and fixed at these values for NPs. The maximum scattering vector Q of the data used for the generation of PDF was 26 Å$^{-1}$. We fit NP data with diffpy-CMI[53] using three different models: periodic models using a single spherical or the lognormal distribution of nanoparticle sizes[70] and a model based on a set of discrete NP structures. The number of refined parameters was kept as low as possible, visual inspection and the residual R$_w$ were used to evaluate the goodness of the fit. The refined parameters for the first two models were the lattice parameter $a$, the U$_{iso}$ of Th and O, the mean particle size (P$_{size}$) and the variance of the lognormal distribution (P$_{sig}^2$). For the model implementing a set of NP structures two parameters were common to all structures, the U$_{iso}$ of Th and O, and two additional parameters were specific to each NP in the set, a scale factor and a lattice expansion parameter. To limit the number of parameters, $\delta_2$, which takes into account the first neighbour interaction, was fixed to 2.0 Å for all fits.

XANES spectra of Th M$_4$ edge were collected in the HERFD mode at the ID26 beamline[71] of the ESRF. The incident energy was selected with a Si(111) double crystal monochromator. Rejection of higher harmonics was achieved with three Si mirrors at 3.0, 3.5 and 4.0 mrad angles relative to the incident beam. The X-ray emission spectrometer[72,73] was equipped with 3 Ge(220) spherically bent crystal analysers (1 m radius) at a Bragg angle of 80° to collect the maximum of the M$_□$ emission line of Th (~3148.6 eV – tabulated value). The energy resolution estimated by the FWHM of the elastic peak was 0.4 eV. Samples were measured as dried powders sealed with double kapton confinement in the sample-holder.

We simulated Th M$_4$ edge XANES with the FDMNES code.[59,60] The scattering potential around the Th absorber was calculated self-consistently within a radius of 5 Å. The best agreement was obtained with the inclusion of a fully screened core-hole and with the Finite Difference Method (FDM). Relativistic effects and spin-orbit interaction were included. An example of the input file used for simulations is provided in ESI.

## Acknowledgements

We acknowledge support from the European Research Council (ERC) under grant agreement N° 759696. T.V.P. acknowledge Russian Science Foundation (grant N° 20-73-00130) for supporting synthesis oand primary characterization of ThO$_2$ nanoparticles. E.G. S.N.K. and K.O.K. acknowledge support by the Russian Ministry of Science and Education under grant N° 075-15-2019-1891. The authors would like to acknowledge the ESRF for providing beamtime. L. A. thanks S. Bauters for fruitful discussions on data analysis.

**Keywords:** actinides • HERFD • HEXS • nanoparticles • ThO$_2$